\documentclass[lettersize,journal]{IEEEtran}
\usepackage{amsmath,amsfonts}
\usepackage{algorithmic}
\usepackage{algorithm}
\usepackage{array}
\usepackage[caption=false,font=normalsize,labelfont=sf,textfont=sf]{subfig}
\usepackage{textcomp}
\usepackage{stfloats}
\usepackage{url}
\usepackage{verbatim}
\usepackage{graphicx}

\usepackage{siunitx}
\usepackage{enumitem}
\DeclareSIUnit\sq{\ensuremath{\Box}}
\usepackage{layouts}
\usepackage{wasysym} 
\usepackage[%
    colorlinks=true,
    pdfborder={0 0 0},
    linkcolor=black,
    citecolor=black,
    urlcolor=blue,
]{hyperref}
\usepackage{cleveref}
\usepackage[style=ieee, maxnames=6, minnames=1]{biblatex}
\crefname{figure}{Fig.}{Figs.}
\Crefname{figure}{Figure}{Figures}
\crefname{equation}{Eq.}{Eqs.}
\Crefname{equation}{Equation}{Equations}
\usepackage{tikz}
\newcommand\submittedtext{%
  \footnotesize This work has been submitted to the IEEE for possible publication. Copyright may be transferred without notice, after which this version may no longer be accessible.}
\newcommand\submittednotice{%
\begin{tikzpicture}[remember picture,overlay]
    \node[anchor=south,yshift=10pt] at (current page.south) {\fbox{\parbox{\dimexpr0.65\textwidth-\fboxsep-\fboxrule\relax}{\submittedtext}}};
\end{tikzpicture}%
}
\usepackage{orcidlink}

\begin{document}

    \title{Dark characterization of Ti/Al LEKIDs for the\\search of axions in the W--band.}
    \author{
        Victor Rollano\orcidlink{0000-0001-6878-2609}, Alejandro Pascual Laguna\orcidlink{0000-0001-7293-4150}, David Rodriguez\orcidlink{0000-0002-0795-7724}, Martino Calvo\orcidlink{0000-0002-8752-6325}, Maria Teresa Magaz\orcidlink{0000-0002-6008-2971},\\Daniel Granados\orcidlink{0000-0001-7708-9080}, Alessandro Monfardini, Alicia Gomez\orcidlink{0000-0002-8752-1401}
        \thanks{Manuscript received September 08, 2025; This work has received support from grants PID2022-137779OB-C41, PID2022-137779OB-C42 and JDC2023-051842-I funded by the Spanish MCIN/AEI/10.13039/501100011033, by the EU “NextGenerationEU”/PRTR and by the “ERDF A way of making Europe”. IMDEA Nanoscience acknowledges financial support from the “Severo Ochoa” Programme for Centres of Excellence in R\&D (CEX2020-001039-S) and CAB from the CSIC Research Platform PTI-001, from “Tecnologías avanzadas para la exploración del Universo y sus componentes” (PR47/21 TAU-CM) and Mad4Space-TEC-2024/TEC-182 projects both funded by Comunidad de Madrid, by “NextGenerationEU”/PRTR.}
        \thanks{ V.~Rollano, A.~Pascual Laguna, D.~Rodriguez,  M.~T.~Magaz, A.~Gomez  are with  Centro de Astrobiología (CSIC-INTA), Torrejón de Ardoz, 28850, Spain. \emph{(Corresponding authors: V.~Rollano, A.~Gomez; email: vrollano@cab.inta-csic.es, agomez@cab.inta-csic.es.)}}
        \thanks{D.~Granados is with the Instituto Madrileño de Estudios Avanzados en Nanociencia (IMDEA Nanociencia), 28049, Madrid, Spain.}
        \thanks{M.~Calvo and A.~Monfardini are with the Institut Néel, CNRS and Université Grenoble Alpes, Grenoble 38042, France.}%
        \thanks{Color versions of one or more figures in this article are available at https://doi.org/nn.nnnn/TTAS.yyyy.nnnnnnn. Digital Object Identifier nn.nnnn/TTAS.yyyy.nnnnnnn}
    }
            
    \maketitle
    \submittednotice
    
    \begin{abstract}
        We report the electrical (dark) characterization of lumped–element kinetic inductance detectors (LEKIDs) fabricated from a Titanium/Aluminum bilayer and designed for broadband absorption in the W--band (75–110 GHz). These detectors are prototypes for future QCD axion search experiments within the Canfranc Axion Detection Experiment (CADEx), which demand sub–$10^{-19}~\text{W}/\sqrt{\text{Hz}}$ sensitivities under low optical backgrounds. We combine a Mattis–Bardeen analysis to the temperature dependence of the detector parameters with noise spectroscopy to determine the electrical noise equivalent power (NEP). The minimum measured value for the electrical NEP is $\sim 3 \times 10^{-19}~\text{W}/\sqrt{\text{Hz}}$. Across the measured temperature range, we find that quasiparticle lifetime deviates from the expected BCS recombination law. Our analysis suggests that non-equilibrium relaxation is governed by spatial inhomogeneities in the superconducting gap and phonon diffusion effects. This work sets the road-map to achieve suitable and ultra-sensitive detectors in the W--band for dark matter axion search experiments.
    \end{abstract}
    
    
    \section{Introduction}

        \IEEEPARstart{M}{icrowave} kinetic inductance detectors (MKIDs) have emerged as a leading technology for large-format, ultralow-noise detection at millimeter and submillimeter wavelengths. Their inherent multiplexability, simple planar fabrication, and straightforward cryogenic integration make them attractive for astrophysical and laboratory instruments targeting background-limited performance. MKIDs are now deployed in several (sub)millimeter cameras and spectrometers, including NIKA2~\cite{NIKA2}, CONCERTO~\cite{CONCERTO}, and DESHIMA~\cite{endo2019deshima}. Recently, MKIDs have been confirmed as the selected type of detectors for the PRIMA space telescope of NASA, ratifying its status as a leading detector technology \cite{Glenn2024prima}.

        Within the MKID family, lumped-element KIDs (LEKIDs) offer flexible impedance and absorber design, advantageous for broadband operation in the W--band (75-110\,GHz). We have developed a prototype array of LEKIDs intended for the Canfranc Axion Detection Experiment (CADEx)~\cite{CADEx}, which aims to search for axion dark-matter signatures (via the Primakoff conversion/decay channel) in the $330$-$460~\mu\text{eV}$ mass range, corresponding to W--band photons. As incoherent (direct) detectors, MKIDs are not limited by the quantum noise like coherent receivers are~\cite{Zmuidzinas2003thermalnoise}. This means that, for a given detection bandwidth $\Delta \nu$, direct detectors like MKIDs can match or even beat the sensitivity of coherent detectors, which makes them extremely attractive for dark matter detection experiments. 

        The figure of merit for a power-integrating direct detector is the noise equivalent power (NEP), defined as the input–referred power spectral density that yields an $\text{SNR}=1$ in a $1~\text{Hz}$ output bandwidth. In most applications, the NEP is measured under illumination, ideally being limited by photon noise (shot and, when relevant, bunching), which can mask the detector’s intrinsic performance. In this sense, CADEx requires a NEP as low as $\sim 10^{-20}~\text{W}/\sqrt{\text{Hz}}$.

        This work focuses on the electrical characterization of the CADEx camera prototype in dark conditions, to determine the device electrical NEP which imposes a lower bound on the optical NEP. The paper is organized as follows: Section II presents the fabricated Ti/Al LEKIDs and reviews the experimental details. In Section~III, the radio-frequency (RF) characterization results and their analysis is presented; describing the quasiparticle electrodynamics and reporting the dark electrical NEP. The paper concludes discussing the results and proposing different improvements for future implementations based in these findings.

        \IEEEpubidadjcol
    \section{Experimental section}

        The device studied in this work consists of nine LEKIDs organized in a $3 \times 3$ array, all of them coupled to a microstrip transmission line that serves as readout line. Each detector has a different resonance frequency, with values far for each other to allow multiplexed readout without cross-talking. The sample has been fabricated on a Ti/Al bilayer deposited onto a $275~\mu\text{m}$-thick high-resistivity ($\rho>1~\text{k}\Omega \cdot \text{cm}$) silicon (Si) substrate. The substrate thickness is chosen to approximate a quarter-wave optical thickness near \(90~\text{GHz}\), at the center of the W–band. Native Si oxide is removed in a \(1\%\) hydrofluoric acid bath; within one minute the wafer is loaded into a confocal sputtering chamber with base pressure of $\sim 10^{-9}$ Torr. The Ti/Al stack consists of a $10~\text{nm}$ Ti base layer capped with a $20~\text{nm}$ Al layer. An additional $200~\text{nm}$ Al film is deposited on the backside of the wafer serving as both a reflective backshort and ground plane. During the metal deposition, the argon pressure is maintained at $10^{-3}$ Torr, reaching a target evaporation rate of $0.22 \text{\AA}/\text{s}$ for Ti and $0.24 \text{\AA}/\text{s}$ for Al. We performed the device patterning on a maskless laser-writer photolithography using AZnLOF 2070 negative photoresist. After developing the resist, the uncovered metal was removed using wet etching (90 seconds in Al80 for Aluminum and 180 seconds in TBR19 for Titanium). Then, the remaining resist is removed using an acetone and isopropanol rinse. Additional information on the detector design can be found elsewhere \cite{rollano2025cross}.
        
        \begin{figure*}[t]
            \centering
            \includegraphics[width=2.0\columnwidth]{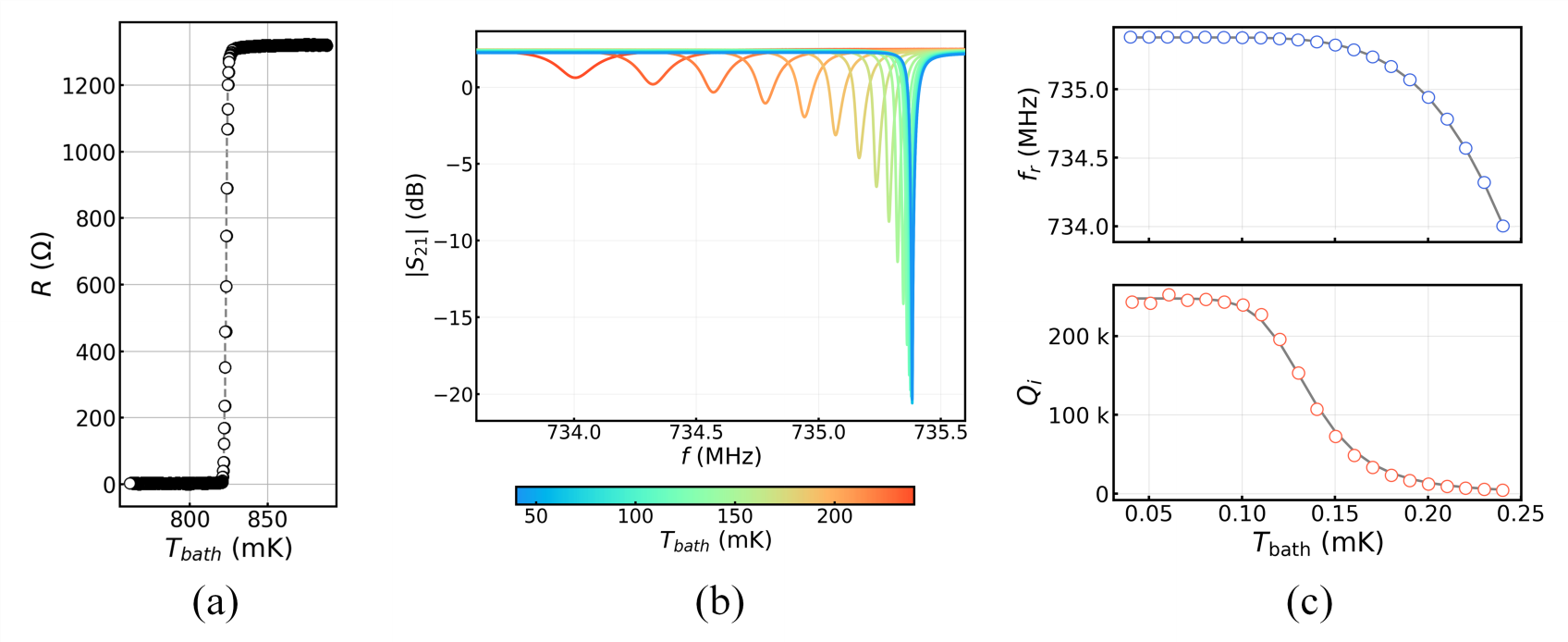}
            \caption{(a) Resistance as a function of temperature for a meander of 10 $\mu$m width and 10.785 mm long measured with a DC current. (b) Transmission amplitude as a function of readout frequency for various bath temperatures between 40 mK and 240 mK. The microwave readout power at the chip level is fixed at –90 dBm. (b) Extracted resonance frequency (top) and internal quality factor (bottom) as functions of bath temperature. The colormap is identical to that used in panel (c). Gray solid lines represent fits of the experimental data using the Mattis–Bardeen model. The Dashed line a is guideline for the eye.}
            \label{Fig1}
        \end{figure*}
        
        The electromagnetic response of the prototype has been characterized in a BlueFors LD250 dilution refrigerator The device is mounted in a closed sample holder in thermal contact with the mixing chamber of the cryostat. The inner walls of the sample holder are painted with carbon-loaded epoxy and SiC grains, an absorber material, trying to maintain the darkest environment as possible. 
        A homodyne IQ readout set up is employed to perform the RF measurements. A microwave signal generator provides a continuous-wave probe that is split by a power divider: one output serves as the local oscillator (LO) for an IQ mixer, while the other passes through a room-temperature variable attenuator (to set the input power) and then into the cryostat. The input coaxial line inside the refrigerator provides $-40~\text{dB}$ of cold attenuation before the signal reaches the device. The transmitted signal is amplified at the $4~\text{K}$ stage by a low-noise amplifier with a noise temperature of $\sim 5~\text{K}$, and fed to the RF port of the IQ mixer for homodyne demodulation using the common LO. The in-phase ($I$) and quadrature ($Q$) outputs are digitized using a National Instruments PXI-5922, operated typically at $1~\text{MSa/s}$ with an effective $22$-bit vertical resolution. 
        
        The choice of using a Ti/Al stack obeys the fact that the superconducting gap $\Delta$ imposes a detection frequency lower limit. Typical MKIDs fabricated from Aluminum cannot detect photons below $2 \Delta_{Al} / h \sim90~\text{GHz}$. Adding a thin layer of titanium below the aluminum tends to lower $\Delta$ due to proximity effect, making the device suitable for the entire W--band \cite{Catalano2015}. As an initial characterization, the superconducting transition was determined by measuring the resistance of a Ti/Al meander as a function of temperature. Figure~\ref{Fig1}(a) shows the DC measurement of the transition. The critical temperature, defined at $0.5R_n$, is $820~\text{mK}$, with $R_n$ being the resistance before the superconducting transition. Using the BCS relation $\Delta_0 = 1.764 k_B T_c$, we obtain a zero-temperature superconducting gap of $\Delta_0 = 124.6~\mu\text{eV}$. In the normal state, the bilayer exhibits a sheet resistance of $R_{s,\square} \approx 1.22~\Omega/\mathrm{sq}$. Within the BCS dirty-limit framework, the kinetic inductance per square can be estimated as $L_{k,\square} = \hbar R_{s, \square} / (\pi \Delta_0)$, yielding $L_{k, \square} \approx 2.02~\text{pH}/\square$.

        \begin{figure}[b!]
            \centering
            \includegraphics[width=1.0\columnwidth]{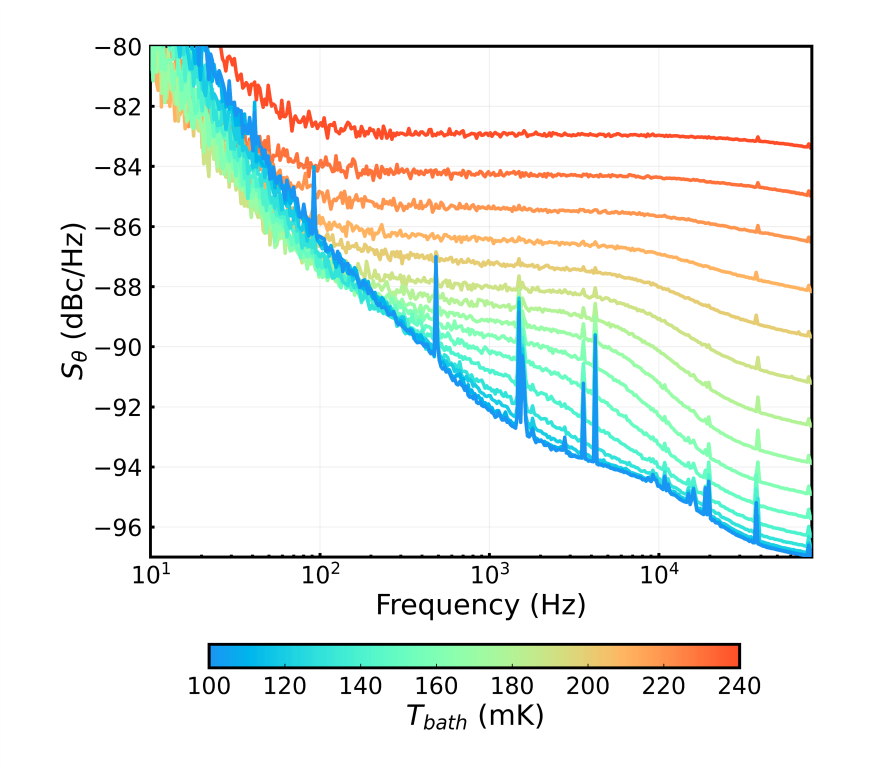}
            \caption{Measured phase noise Power Spectral Density as a function of audio frequency for a single detector measured at different bath temperatures between 40 mK and 240 mK, increasing from bottom to top. Microwave readout power is -90 dBm, while the frequency of the readout signal is fixed at the resonance frequency obtained from fitting each trace. }
            \label{Fig2}
        \end{figure}
        
    \section{Results}
    
        \begingroup\centering
        \subsection{Superconducting electrodynamics}
        \endgroup
        
            We investigated the electrodynamics of the Ti/Al bilayer under dark, thermal–equilibrium conditions by monitoring the detector response versus bath temperature $(T_{bath})$. Figure~\ref{Fig1}(b) shows the variation in the amplitude of transmission parameter $|S_{21}|$ as $T_{bath}$ increases from 40 mK to 240 mK for one representative detector. The resonator was probed with a microwave readout signal of $-90$ dBm at the chip input (i.e., after accounting for the attenuation of the input line). This choice for the readout power is justified as it is the highest value that avoids non-linear response. As the temperature rises, the resonance dip shifts to lower frequencies and broadens, reflecting the increase in kinetic inductance and in the losses of the Ti/Al bilayer.
            
            The transmission data at each temperature were analyzed by fitting to the model described in \cite{probst2015efficient}. From this analysis, we extracted the resonance frequency ($f_r$) and the internal quality factor ($Q_i$) as functions of temperature, shown in Fig.~\ref{Fig1}(c). The same panel also includes fits to the Mattis–Bardeen (MB) model at low temperature, expressed as \cite{Gao2008}:
    
            \begin{equation}
                f_r(T) = f_r(0) \left[ 1 + \frac{\alpha}{2} \left( 1 - \frac{S_2(0)}{S_2(T)} \right) \right],
                \label{eq1}
            \end{equation}
            
            and
            
            \begin{equation}
                Q_i^{-1}(T) = Q_{sat}^{-1} + \frac{2 \alpha}{\pi} \frac{\Delta_0}{\Delta(T)} \frac{S_1(T) S_2(0)}{S_2(T)},
                \label{eq2}
            \end{equation}

            using $S_1(T)$ and $S_2(T)$ as defined in \cite{gao2008equivalence}.
                
            In this model, the kinetic-inductance fraction is defined as
    
            \begin{equation}
                \alpha \equiv \frac{L_k}{L_k + L_g},
                \label{eq3}
            \end{equation}

            where $L_g$ is the geometric inductance of the resonator. The parameters $f_r(0)$ and $Q_{sat}$ are the resonance frequency and the residual value of $Q_i$ in the zero-temperature limit, where MB quasiparticle loss can be considered negligible. For temperatures well below $T_c$, the superconducting gap $\Delta$ is assumed to vary with temperature as \cite{Gao2008}:
            
            \begin{equation}
                \Delta(T) = \Delta_0 \left[ 1 - \sqrt{\frac{2 \pi k_B T}{\Delta_0}} e^{- \Delta_0 / k_B T} \right].
                \label{eq4}
            \end{equation}
           
            The fit to the resonance frequency as a function of temperature, shown in Fig.~\ref{Fig1}(c), yields $\alpha = 0.34$, $\Delta_0 = 123.72~\mu$eV (corresponding to $T_c \approx 814.7~\text{mK}$), and $f_r(0) = 735.38$ MHz. The $\Delta_0$ value obtained from both MB analysis and DC transport measurements are compatible, ensuring sensitivity to excitations above $\Delta_{\mathrm{Ti/Al}}/h \sim 59~\text{GHz}$, making this material suitable for detection in the entire W-band ($75~\text{GHz}-110~\text{GHz}$). Best-fit parameters for $Q_i^{-1}(T)$ are $\alpha = 0.374$ and $Q_{sat} = 2.45 \times 10^5$, with the value $\Delta_0$ being fixed from the resonance frequency fit. 
            
            We have compared the values obtained for $\alpha$ from the fitting analysis with the ones that can be computed from electromagnetic simulations of the resonator design using Sonnet \cite{SONNET}. The geometric inductance obtained from EM simulations is $L_g \approx 33~\text{nH}$. For an inductor of $\square = 9256 ~ \text{sq}$, the total kinetic inducatnce is $L_k \approx 19 ~\text{nH}$, yielding a kinetic fraction of $\alpha_{sim} \approx 0.37$.

        \begingroup\centering
        \subsection{Noise spectra and quasiparticle dynamics}
        \endgroup
            In this section, we present noise spectra analysis as a starting point to later characterize the sensitivity of the detectors. We acquired noise traces as a function of the temperature of the quasiparticle bath $T_{bath}$. Each trace is composed of 20 noise samples of $1~\text{s}$ sampled at $500~\text{kSa/s}$. The microwave stimulus is set in the resonance frequency of the resonator at each value of $T_{bath}$. The readout power, as in the previous section, is mantained at $-90~\text{dBm}$.

            \begin{figure}[t]
                \centering
                \includegraphics[width=1.0\columnwidth]{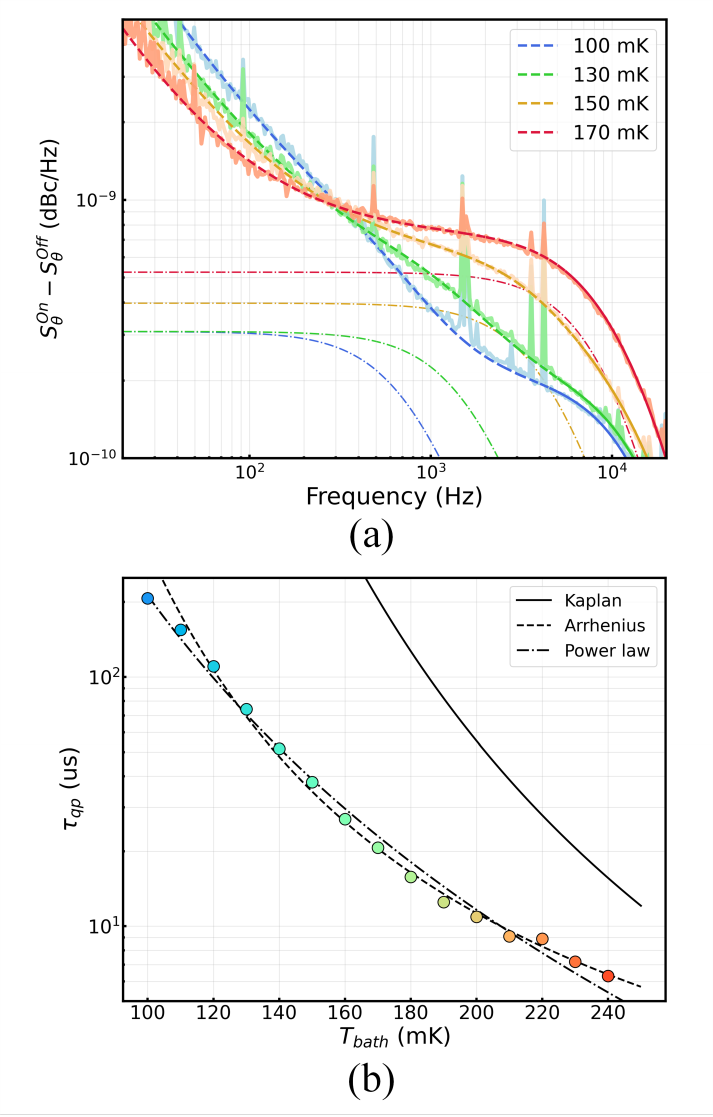}
                \caption{(a) Fits of measured noise power spectral densities together with their corresponding roll-off components, shown for four representative bath temperatures (100 mK, 130 mK, 150 mK, and 170 mK). The fitting is performed using the model described in Eq. \ref{eq5}. Dashed lines indicate the full model fit, while dash–dot lines highlight the quasiparticle Lorentzian roll-off contribution. PSD curves are shown after subtraction of the off-resonance spectra, thereby removing the amplifier noise floor. (b) Quasiparticle recombination time as a function of bath temperature, extracted from the Lorentzian fits in the PSDs. Black solid, dashed and dash-dotted curves represent the Kaplan, modified Arrhenius and phonon-electron coupling models.}
                \label{Fig3}
            \end{figure}
            
            The phase noise single-sided power spectral density $S_{\theta} (f)$ measured for several bath temperature values is shown in Fig~\ref{Fig2}. At low temperatures, we observe a noise spectrum that presents two different roll-offs. The high frequency one corresponds to the resonator dynamics, with a decay rate $\kappa = 20.22~\text{kHz}$, given by the expression $\kappa = f_r / (2 Q_L)$. Stepping down in frequency, we find a second roll-off, which we attribute to the quasiparticle generation-recombination (GR) processes. priorAs temperature increases, the GR bandwidth upshifts towards higher frequencies and increases its noise level, a behavior expected from the GR kinetics\cite{deVisser2014fluctuations}. At high temperatures, the GR roll-off hides the resonator bandwidth completely. Finally, the high-frequency band is dominated by the white noise introduced by the amplifier, while at low frequencies the spectrum is dominated by a 1 over f slope.
    
            \begin{figure*}[t]
                \centering
                \includegraphics[width=2.0\columnwidth]{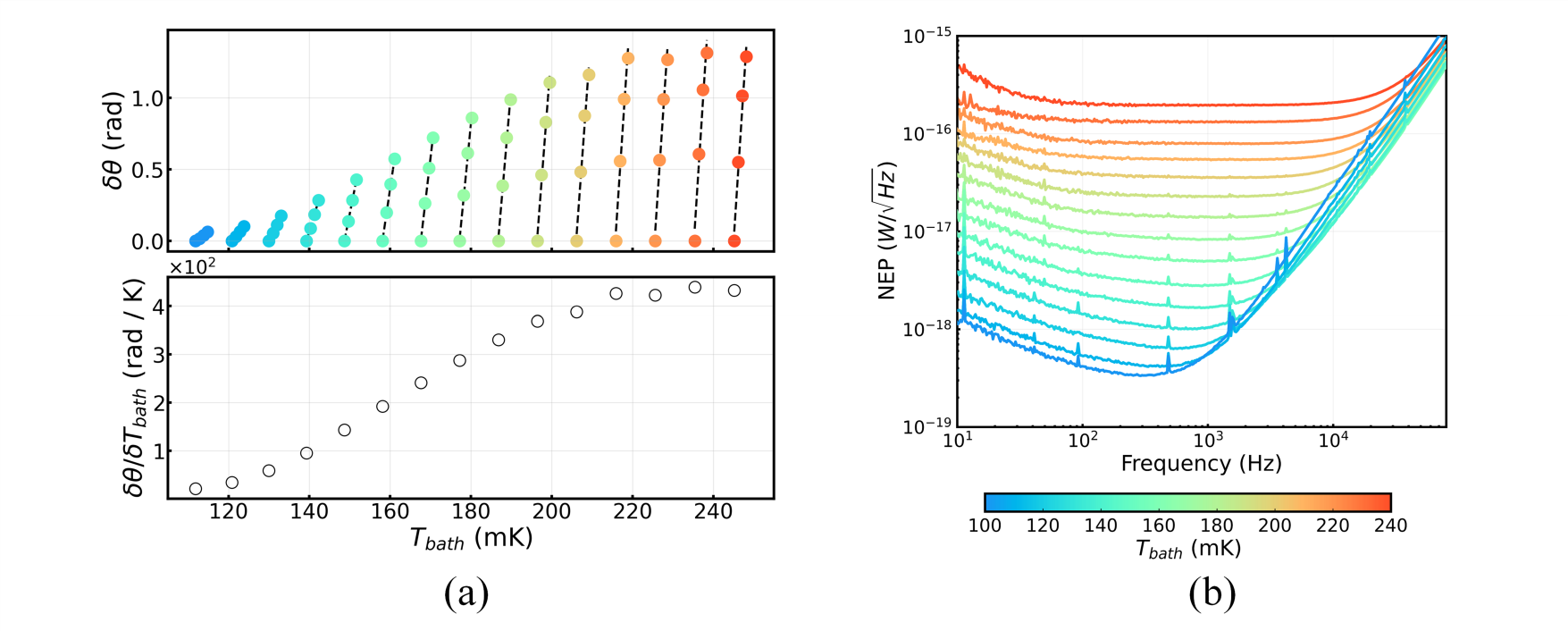}
                \caption{(a) Phase response (top) and phase responsivity (bottom) as functions of bath temperature. Each phase response is referenced to the center of the corresponding resonance trace measured at the same temperature. (b) Measured electrical noise-equivalent power (NEP) in the resonator phase as a function of audio frequency for bath temperatures between 100 mK and 240 mK.}
                \label{Fig4}
            \end{figure*} 
            
            The phase noise single-sided PSD due to the GR dynamics can be modeled as:
    
            \begin{equation}
                S_{\theta}(f) = \frac{4 N_{qp} \tau_{qp}}{1 + (2 \pi f \tau_{qp})^2} \left( \frac{\delta \theta}{\delta N_{qp}}\right)^2
                \label{eq5}
            \end{equation}
    
            where $\tau_{qp}$ is the GR characteristic time and $N_{qp}$ the mean quasiparticle number.
           
            In Fig.~\ref{Fig3}(a) we show best-fit results for four representative phase noise spectra measured at $100,\mathrm{mK}$, $130,\mathrm{mK}$, $150,\mathrm{mK}$, and $170,\mathrm{mK}$. Each on-resonance spectrum has been baseline-corrected by subtracting the off-resonance spectrum, thereby removing the amplifier white-noise floor contribution. The model used to fit the phase noise consists on the Lorentzian shape from Eq.~\ref{eq5}, with the addition of a 1/f contribution and another Lorentzian shape due to the resonator roll-off, with a characteristic time $\tau_{res} = 2 \pi \kappa^{-1}$. The dash-dot curve in Fig.~\ref{Fig3}(a) represents the isolated GR component given by Eq.~\ref{eq5}.

            Figure \ref{Fig3}(b) depicts the obtained GR time as a function of temperature $\tau_{qp}\left(T\right)$. We first compare the extracted data against the Kaplan quasiparticle recombination model \cite{kaplan1976quasiparticle},
    
            \begin{equation}
                \tau_{qp} (T) = \frac{\tau_0}{\sqrt{\pi}} \left[ \frac{k_B T_c}{2 \Delta(T)} \right]^{5/2} \sqrt{\frac{T_c}{T}}e^{\Delta_0 / k_B T},
                 \label{eq6}
            \end{equation}
    
            using $\Delta_0$ fixed by the measured critical temperature and the MB fits to $f_r(T)$ and $Q_i(T)$. With these inputs, the Kaplan model does not describe the GR dynamics in the Ti/Al bilayer (see black solid line), even when using a phonon trapping factor $\tau_{qp}^* = \tau_{qp} (1 + \beta)$ \cite{Rooij2021} that modifies the apparent quasiparticle lifetime. The gap needed to reproduce the kink observed in the measured $\tau_{qp}$ is too low to be fitted by the proximitized gap obtained from the measured superconducting transition. This may be indicative of some other processes affecting the quasiparticle kinetics in the detector and modifying the temperature dependence. 
            
            To extract physical information from these data, we have used two different phenomenological models:
            \begin{enumerate}[leftmargin=0.4cm]
                \item[i.] We fitted the GR recombination time with a modified Arrhenius model 
                \begin{equation}
                    \tau_{qp} = \tau_0 \exp{\left[ m \Delta(T) / k_B T\right]}.
                    \label{eq7}
                \end{equation}
                The fit yields $m = 0.47$ and $\tau_0 = 388~\text{ns}$. The result $m < 1$ can be interpreted as an effective activation energy below $\Delta_0$ due to subgap states or gap spatial inhomogeneity associated with disorder or proximity effects in the bilayer \cite{gao2012titanium, bueno2014anomalous}. The resulting fit is plotted in Fig. \ref{Fig3}(b) using a dashed black curve.
                \item[ii.] Exploring the possibility that the dominant energy dissipation mechanism is the electron-phonon coupling, which introduces a heavy dependence of $\tau_{qp}$ on temperature, we have also used a power-law to fit the data as \cite{sergeev2000electron, kardakova2013electron}
                \begin{equation}
                    \tau_{qp} = A T ^{-n}.
                    \label{eq8}
                \end{equation}
                The fit using this model is shown in Fig.~\ref{Fig3}(b) with a dash-dot black line. The fit yields $n \approx 4.1 $, typical of disordered superconductors \cite{sidorova2020electron}, and $A \approx 1.37 \times 10^{-8} K\cdot s^{-4.1}$.
            \end{enumerate}

        \begingroup\centering
        \subsection{Dark sensitivity}
        \endgroup
        
            The phase response of the detector as a function of $T_{bath}$ is presented in the top panel of Fig.~\ref{Fig4}(a). At each temperature, the complex transmission was measured with a 1 ms microwave readout pulse at constant temperature, averaged over ten repetitions. As in previous sections, the microwave power is maintain at $-90 ~\text{dBm}$. Then, bath temperature is increased 1 mK and the complex trace is acquired again following the same procedure. Using four response points around each main bath temperature (stepped 1 mK), we extract the electrical phase responsivity $\delta \theta / \delta T_{bath}$ as the slope of the phase response, as pictured in Fig.~\ref{Fig4}(a). The obtained electrical phase responsivity is plotted in the bottom panel of Fig.~\ref{Fig4}(a). 
            
            From this, we obtain the dark electrical responsivity as:
    
            \begin{equation}
                \frac{\delta \theta}{\delta P_{dark}} = \frac{\eta_{pb} \tau_{qp} (T) }{\Delta(T)} \frac{\delta \theta}{\delta N_{qp}},
                 \label{eq9}
            \end{equation}
    
            where $\eta_{pb}$ is the pair-breaking efficiency and $N_{qp}$ the mean quasiparticle number in thermal equilibrium. We related $\delta \theta/\delta N_{qp}$ to the measured $\delta \theta / \delta T_{bath}$ using
    
            \begin{equation}
                \frac{\delta \theta}{\delta N_{qp}} = \frac{\delta \theta}{\delta T_{bath}} \left(\frac{\delta N_{qp}}{\delta T_{bath}}\right)^{-1},
                 \label{eq10}
            \end{equation}
    
            with $N_{qp}(T)$ given, for $T \ll T_c$, by the standard BCS low-temperature asymptotic expression
            
            \begin{equation}
                N_{qp} (T) = 2 N_0 V \sqrt{2 \pi k_B T \Delta(T)} e^{-\Delta(T) / k_B T},
                 \label{eq11}
            \end{equation}
            
            where $N_0$ is the single–spin electronic density of states at the Fermi level and $V$ is the active superconducting volume (i.e., the inductor volume). We approximated the effective $N_0$ of the bilayer as the weighted average
            
            \begin{equation}
                N_0^{eff} = \frac{N_0^{Al} t_{Al} + N_0^{Ti} t_{Ti}}{t_{Al} + t_{Ti}},
                \label{eq12}
            \end{equation}
            
            where $t_{Al}$ and $t_{Ti}$ represent the thickness of the Aluminum and Titanium layers respectively, and with $N_0^{Al} = 1.74 \times 10^{10}~\mu \text{m}^{-3} \text{eV}^{-1}$ and $N_0^{Ti} = 4.1 \times 10^{10}~\mu \text{m}^{-3} \text{eV}^{-1}$ \cite{zhao2017exploring}.

            We use $\Delta(T)$ obtained from the MB analysis of $f_r(T)$ and $Q_i(T)$ because the dark electrical responsivity is an equilibrium quantity that depends on the thermal quasiparticle population $N_{qp}$. The MB fit yields a self-consistent $\Delta(T)$ tied to the film’s equilibrium electrodynamics. In contrast, the $\tau_{qp}$ analysis probes dynamics (small fluctuations $\delta N$ around $N_{qp}$) and can be influenced by nonequilibrium effects such as readout power, phonon bottlenecking, or trapping. Using the MB-derived $\Delta(T)$ to compute $N_{qp}(T)$ cleanly separates the equilibrium DOS (which sets the mean $N_{qp}$) from the relaxation kinetics (which set $\tau_{qp}$), avoiding biases from nonequilibrium corrections in the lifetime fit.
    
            Finally, we obtain the dark electrical NEP as
    
            \begin{equation}
                NEP_{dark} ( f )= \sqrt{S_{\theta}}\left(\frac{\delta \theta}{\delta P_{dark}}\right)^{-1} \sqrt{ 1 + ( 2 \pi f \tau_{qp} )^2}~~.
                \label{eq13}
            \end{equation}
    
            The measured dark electrical NEP as a function of frequency for several values of the bath temperature from 100 mK to 240 mK is shown in Fig. \ref{Fig4}(b). The NEPs are determined from the measured noise spectra and responsivities. We find that the best value of the electrical NEP, for a frequency of $200~\text{Hz}$ and $T_{bath} = 100 ~\text{mK}$ is $3 \times 10^{-19} ~\text{W} / \sqrt{\text{Hz}}$. 
    \\
    
    \section{Discussion}
    
        Our work shows that the Ti/Al bilayer low temperature physics in thermal equilibrium are well captured by the MB framework. The analysis of the $f_r(T)$ and $Q_i(T)$ yield a superconducting gap of $\Delta_0 \approx 124 ~\mu\text{eV}$, a value consistent with the measured $T_c$. However, the quasiparticle recombination time extracted from GR roll-off in the PSDs deviates from the typical Kaplan model over our temperature range. Quasiparticle recombination in the Ti/Al stack following a non-typical law suggests that other mechanisms rather than pure BCS recombination set the practical sensitivity of the detector. Our analysis using two different phenomenological models (i.e., the modified Arrhenius form and a power law) reproduce the measured data. The first indicates that an effective activation energy bellow $\Delta_0$ is in play (since the phenomenological parameter $m$ is lower than one), while the second points that electron-phonon processes dominate the quasiparticle dynamics in the system.

        Both interpretations are natural in a Ti/Al bilayer. Proximity effects \cite{hosseinkhani2018proximity} and material inhomogeneities \cite{bueno2014anomalous} can produce sub-gap states that dominate the GR kinetics, yielding an apparent $m < 1$ in the modified Arrhenius model. On the other hand, the quasiparticle dynamics can be influenced also from phonon processes like for example phonon trapping or diffusion bottlenecks originated in thermal phonons that cannot dissipate due to acoustic mismatch between layers \cite{kaplan1979acoustic,rostem2018enhanced}. These phonon processes tend to steepen the temperature dependence toward $T^{-n}$ with $n \sim 4$. In our case, gap inhomogeneity and phonon transport dominate the GR dynamics, following previous studies based on disordered superconductors like TiN \cite{bueno2014anomalous}. 

        Combining measured responsivities with the noise spectra, we determine a minimum dark electrical NEP of $\sim 3 \times 10^{-19} ~\text{W}\sqrt{\text{Hz}}$ at 200 Hz and $T_{bath} = 100~\text{mK}$, which sets a lower bound on the optical NEP for these devices. 
        We propose different engineering strategies that can be implemented to improve the NEP value: (i) using membranes or phononic structures as substrate can enhance phonon escape and suppress re-breaking, improving responsivity and thus the NEP value \cite{Rooij2021, deVisser2021phonon}; (ii) reducing the volume of the absorber to increase responsivity \cite{deRooij2025volume}; and (iii) improving thermalization in our set-up will also improve phonon dissipation, avoiding quasiparticle poisoning from phonon diffusion mechanisms.

    \section{Conclusions}
        The presented results constitute a complete study of the superconducting electrodynamics of Ti/Al LEKID aimed at axion search and astrophysical observations in the W--band (75-110 GHz). Using a Mattis-Bardeen analysis we obtained a thermal equilibrium description of the Ti/Al bilayer, consistent with the measured $T_c$ and EM estimations, confirming suitability across the full W--band. From phase-noise spectroscopy, we find that quasiparticle recombination kinetics deviate from BCS prediction; in contrast, the data are well captured by phenomenological models, pointing to gap inhomogeneity and phonon difussion bottlenecks as dominant non-equilibrium processes. 

        In summary, these results establish a starting point for the development of dark matter axion detectors based on Ti/Al LEKIDs. The developed prototype has achieved competitive sensitivity under dark conditions. Although the measured NEP sets a lower bound, it still remains uncertain whether the observed deviation from the expected BCS recombination dynamics will affect the optical NEP. Future work will address this question through dedicated experiments under controlled illumination.        

    \section*{Acknowledgments}
        
        The authors would like to thank R.~Ferrándiz from INTA (Spain) for his support designing and fabricating the chip holder. We would like to thank also the CADEx Collaboration for useful discussions.

    \printbibliography
    
\end{document}